\documentclass[twocolumn,showpacs,preprintnumbers,amsmath,amssymb]{revtex4}

\usepackage{graphicx}
\usepackage{dcolumn}
\usepackage{bm}

\begin{document}

\preprint{APS/123-QED}

\title{Particle deposition after droplet evaporation on super-hydrophobic micro-textured surfaces} 

\author{P. Brunet}
\email{philippe.brunet@univ-paris-diderot.fr}
\affiliation{Laboratoire Mati\`ere et Syst\`emes Complexes UMR CNRS 7057, Batiment Condorcet, 10 rue Alice Domont et L\'eonie Duquet 75205 Paris Cedex 13, France}
\date{\today}

\begin{abstract}

We study the size and shape of the final deposit obtained when a drop with colloidal particles has dried on a super-hydrophobic surface made of micro-posts. As expected, most of the particles lie inside a circular area, which radius roughly corresponds to the Laplace pressure threshold for liquid impalement inside the structure (Cassie-Wenzel transition), inducing a coffee-stain deposit due to contact-line pinning. Less expected is the observation of tiny deposits on top of posts in the area external to the main ring, despite the low macroscopic liquid/solid friction. Experiments are carried out varying the concentration in particles and initial volume of drops, in order to determine the influence of these parameters on the size distribution of deposits. A microscopic insight of the tiny deposits is proposed, based on recent experiments of non-volatile liquid sliding drops.

\end{abstract}


\maketitle                              

\section{Introduction}

Free-evaporation of particle-laden drops is a commonly utilized technique to deposit material on a surface, although this apparently simple process involves complex and not yet well-understood phenomena \cite{Plawsky08}. Evaporation leads to triple-line receding, coupled to heat exchange with the surrounding vapor and the substrate. Especially tricky is to predict what occurs at the vicinity of the drop triple-line. More than a decade ago, Deegan \textit{et al.}\cite{Deegan97} evidenced that the interplay between contact-line pinning and evaporation would result in the clustering of colloidal particles near the contact-line: particles are pushed outwards to the periphery as a consequence of a diverging evaporating flux at the contact-line. This process leads in turn to the so-called "coffee-stains"\cite{Deegan97}, including in living suspensions \cite{Nellimoottil07,Baughman10}. The formation of this ring of particles is sometimes undesirable, and can be prevented by using a volatile solvent \cite{Hu_Larson06} and/or a conductive substrate \cite{Ristenpart07}. However, drying-induced deposition of particles can be induced on purpose to generate regular patterned deposits at the trailing edge of a receding contact-line \cite{Rio2006,Bodiguel2010,Berteloot12}. The general idea is to use the huge capillary forces involved during drying to provoke the self-assembling of particles, which has numerous applied prospectives like for instance in the production of nano-crystals \cite{Narayanan04} or in the analysis of bio-samples by mass-spectroscopy \cite{Ressine08} .
Despite many recent studies in this field, no definitive criteria could be found to predict the subtle coupling between evaporation, deposition and contact-line pinning and depinning cycles \cite{Sangani09}. One of the major challenge is to overcome these complexities in order to be able to control the size and location of final deposits. 

\begin{figure}[h!]
\begin{center}
\includegraphics[scale=0.38]{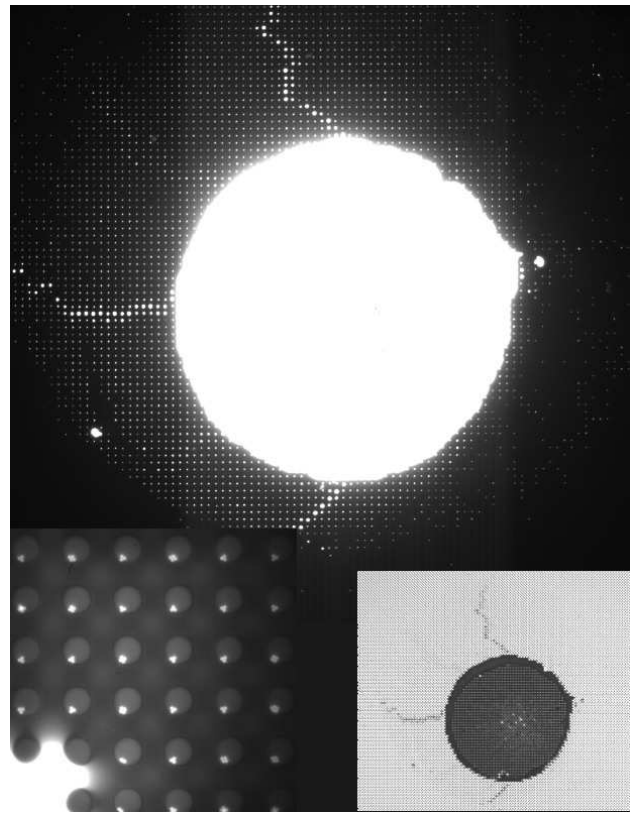}
\caption{The shape of a deposit of colloidal particles after the evaporation of a 10 $\mu$l water drop on a superhydrophobic surface made of micro-pillars. Using fluorescence lighting, particles clearly appear as numerous tiny white spots on top of micro-pillars, additionally to a large circular central deposit. \textit{Left insert -} Zoomed view at the vicinity of the top-right area of the main large deposit. \textit{Right insert -} The same deposit observed with white light.}
\label{fig:depot_large}
\end{center}
\end{figure}

To use a micro-textured substrate would in principle bring an additional level of complexity. However, in the specific case of highly water-repelent "super-hydrophobic" (SH) surfaces, which properties result from an interplay between micro- and/or nano-texturation and low-surface energy coating \cite{Onda96}, it is expected that the final deposit should have a simpler shape than for a flat surface. The reason holds in that on such surfaces, the drop sits on top of the texture elements (the Cassie-Baxter state \cite{Cassie,Bico99}), leading to small liquid/solid contact surface and subsequently to weak pinning forces. 

In this paper we show that, despite weak pinning, such liquid repellent surfaces can lead anyway to particles deposition, in a very accurate and specific manner. We present experiments of water drops containing spherical, monodisperse colloidal particles, drying on model SH surfaces made of straight cylindrical micro-pillars. During evaporation, the drop recedes at almost constant contact-angle \cite{McHale05,Tsai10} and the weak pinning - together with high apparent contact angle - would in principle prevent any particle deposition near the contact-line (the divergence of the evaporating flux requires pinning \cite{Deegan97}). As a consequence, one has considered such surfaces as being anti-fouling or self-cleaning due to that water drops sliding on the surface should grab any dirt from it, and that no particle in suspension inside the drop should be redeposited. In the realm of aquatic plants or insects, some natural surfaces use this property to keep themselves dry and clean \cite{Barthlott97,GaoJiang04}.

Nevertheless, SH or super-oleophobic (SO) surfaces suffer from limitations in their highly non-wetting properties, a major one being that most of them are unable to repel liquid if the pressure in this liquid overcomes a given threshold value, for instance if the liquid impacts on the surface with large enough velocity \cite{Bartolo06,Reyssat06,Lapierre10,Nguyen10}. In evaporation experiments, internal pressure gradually increases due to capillary pressure that scales like the inverse of the drop radius: $P_c = \frac{2 \gamma}{R_c} $. Therefore, the liquid should penetrate into the texture (the "Wenzel state" \cite{Wenzel,Herminghaus00}) once the drop radius is small enough \cite{McHale05,Bartolo06,Reyssat08,Tsai10}, leading to strong pinning. In turn, particles deposition onto and within a ring should occur. Figure \ref{fig:depot_large} shows that the deposit is located in a rather circular area, smaller than that of the initial drop. This area corresponds to the critical radius $R_c$. However, more careful observations reveal that tiny amounts of particles have also been deposited on top of pillars. This is clearly visible in the bottom-left inset of Fig.~\ref{fig:depot_large}. Therefore, these deposits appeared before the drop entered the Wenzel state. The amount of particles on pillars decreases with the distance to the ring, and for the case of one micron-sized particles, it is possible to deposit down to one or two particles per pillar. This phenomenon opens the possibility to sort and extract individual particles, which has many practical interests: for instance in biological cells sorting or in catalysis localization in epitaxial growth. This has also important consequences in evaporation-induced pre-concentration of cells suspensions \cite{Neto_etal11,Shao_etal12}.



In the present paper, we present observations and quantitative measurements on the distribution of the deposits. From these results obtained with various initial conditions, we propose a tentative explanation for the size and spatial distribution of the on-posts deposits. We relate these deposits to recent observations of drops sliding on SH surfaces, which under some conditions leave a trail of tiny droplets on each post.

\section{Experimental setup}

The setup is fairly simple, as it consists in letting a drop of water seeded with particles, to evaporate on a micro-textured surface. The temperature and relative humidity (RH) rate were measured during all experiments, and were found to range respectively between 21 C and 25 C, and between 35 \% and 40\% of RH.

\begin{figure}
\begin{center}
\includegraphics[scale=0.09]{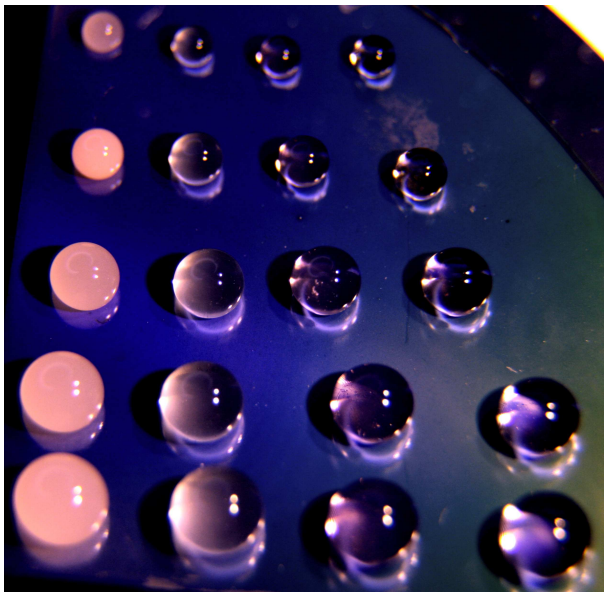}

\includegraphics[scale=0.2]{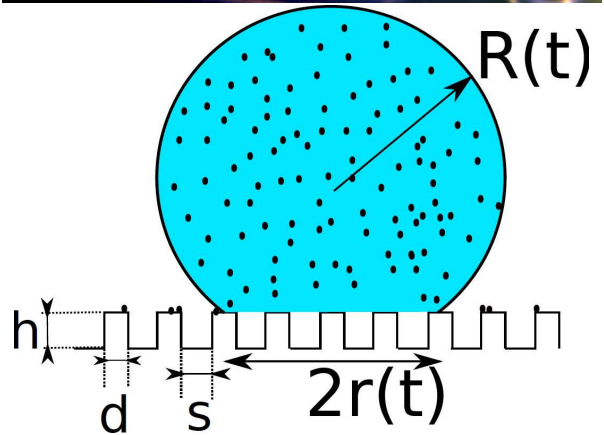}
\caption{\textit{Up} - Initial shape of particle-laden drops on SH surfaces made of micro-posts (see text for details), of various volume (3 to 20 $\mu$l), with two different concentrations (0.01~\% and 0.1~\%) and two particle diameters (50 nm and 1 $\mu$m). \textit{Down} - Scheme of the droplet with relevant lengths.}
\label{fig:image_gouttes_bleues}
\end{center}
\end{figure}

The starting stage of each experiment is depicted in Fig. \ref{fig:image_gouttes_bleues}-\textit{Up}. Each drop initially lies in a Cassie state, with radius $R$ and base radius $r$, both evolving with time (see Fig.~\ref{fig:image_gouttes_bleues}-\textit{Down}). The drop being suspended on top of pillars, the contact-line recedes and jumps inwards from pillars to pillars as liquid evaporates. Rarely, a drop contact-line gets pinned before receding - probably due to that the liquid deposition was not soft enough, and in this case the experiment is rejected. During evaporation phase, the surface and liquid drops were protected by a glass shield cover, to prevent from dirt, dust or air stream to perturb the experiments.

Using a 1-20 $\mu$l micropipette (Nichiro) to take a sample of suspension, carefully mixed and prepared with a 0.1 mg-accuracy balance (Mettler-Toledo), the drop volume $V_0$ and particle initial concentration $c_0$ are well controlled. In practice, we varied $V_0$ from 2 to 30 microlitres, and $c$ from 0.005 \% to 1\% in mass. Particles are spherical and made of latex (Thermoscientific). We used particles of diameter $d_p$=10 $\mu$m, 1 $\mu$m and 50 nm, although the results presented in this paper mainly concern the two latter diameters. 

\begin{figure}
\begin{center}
\includegraphics[scale=0.38]{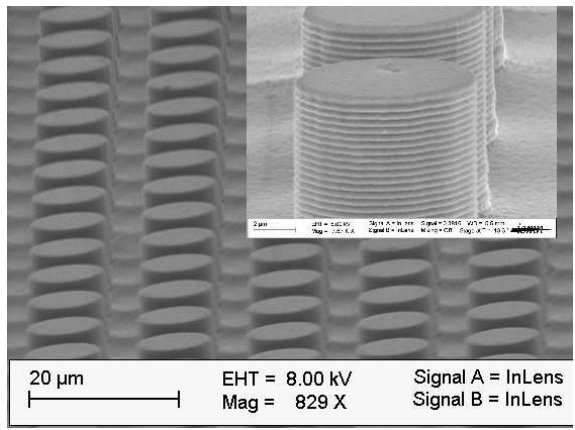}
\caption{SEM image of the micro-textured surface made of cylindrical posts of height $h$ = 10 $\mu$m, diameter $d$ = 10 $\mu$m, and interspace $s$ = 5 $\mu$m. The inset on the top-right reveals a magnified view of a post. The striations are due to the ion etching process that combines successive cycles of etching and passivation.}
\label{fig:piliers_10_5_10}
\end{center}
\end{figure}

Textured surfaces are made from 3-inch silicon wafers, on which standard optical lithography and deep reactive ion etching (Silicon Technology System) are operated. The details of the process is given elsewhere \cite{Lapierre10}. The textured surface is then oxidized by O$_2$ plasma and soaked for 6 hours in a bath in which a low surface-energy liquid Octadecyl-trichlorosiloxane (OTS) dissolved at 1 \% volume in n-hexane. As a result, a monomolecular layer coats the whole texture, making the surface super-hydrophobic.
The final result is depicted in Fig. \ref{fig:piliers_10_5_10}. Although experiments were attempted varying geometrical parameters of the texture, most results presented in this paper were obtained with surfaces made of cylindrical pillars of height $h$ = 10 $\mu$m, diameter $d$ = 10 $\mu$m and interspacing $s$ = 5 $\mu$m. The non-wetting properties are inferred by direct observation that any water drop on the surface appears spherical (see Fig~\ref{fig:image_gouttes_bleues}). Careful contact-angle measurements yield to $\theta_a$ = 140$^{\circ}$ and $\theta_r$ = 125$^{\circ}$, with an accuracy of 2$^{\circ}$.

The images of the final deposits are acquired with a high resolution (2048$\times$2048) Hamamatsu camera C9300-024, connected to a microscope (Olympus) with lenses from 4X to 150X. The deposits are observed either with white light, either with color filters using low light as particles are treated with a fluorescent coating. This fluorescence allowed us to obtain images with high contrast revealing any single particle (see Fig.~\ref{fig:depot_large}) especially for 1$\mu$m particles. 

\section{Qualitative description of the deposits after evaporation}

In this section, we give qualitative insights on the general shape of deposits of particles left on the micro-structure. Figure \ref{fig:depot_large} clearly shows the presence of a main central deposit of rather circular shape, surrounded by many tiny deposits on each posts. Considering that the on-posts particles were deposited while the drop was in a Cassie state, let us define the ensemble of these deposits as "Cassie deposits". Similarly, the central deposit consisting of particles on and between posts is denoted as the "Wenzel deposit" in the following. 

A rough observation of Fig.~\ref{fig:depot_large} reveals that each of these deposits has a specific size, depending on various parameters like radial, distance to the main deposit, orientation angle with respect to the texture main axes, size of particles, initial concentration and drop volume, ... Therefore, we first try to isolate the different influences of these parameters. 

\subsection{Drops of large particles suspensions}

\begin{figure*}
\begin{center}
\includegraphics[scale=0.35]{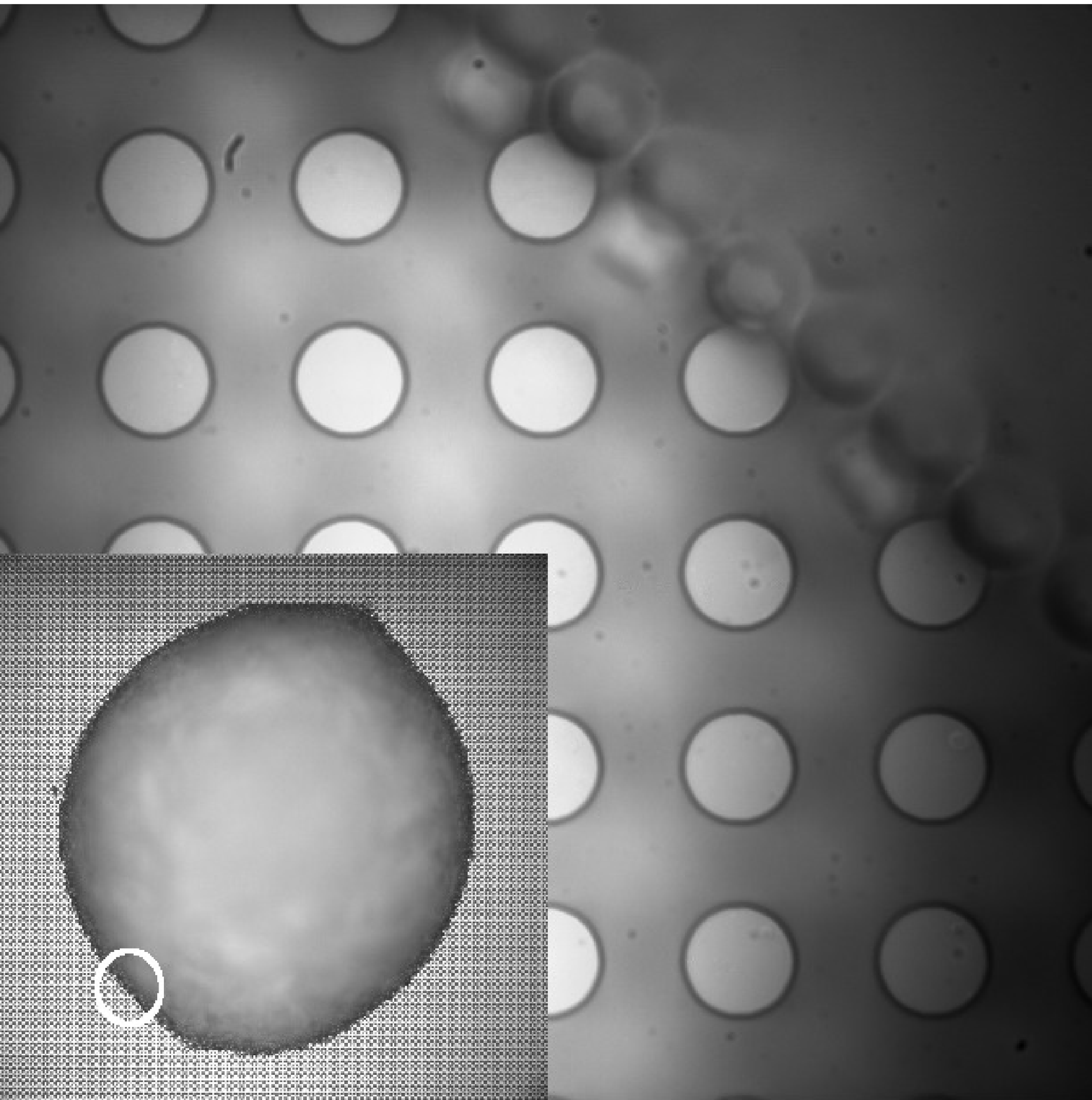}
\includegraphics[scale=0.35]{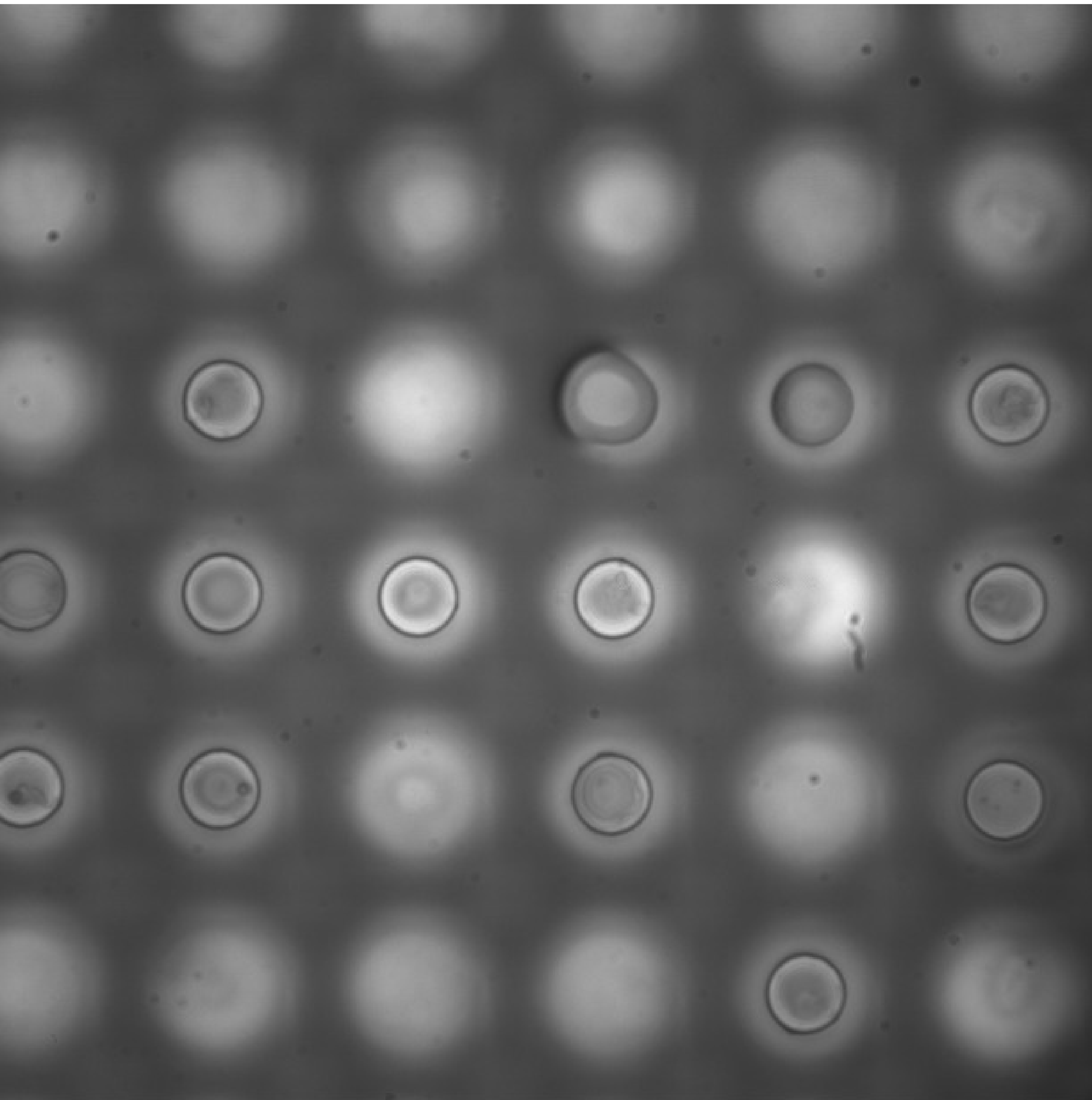}
\caption{\textit{Left } - An extract at the edge of a deposit after the evaporation of a 10 $\mu$l drop of a suspension of 10 $\mu$m particles at initial concentration $c_0$ = 1 \%. The left-bottom insert shows the whole deposit, the white circle representing the vicinity of the extracted area. \textit{Right} - Smaller yeast cells remaining on top of posts after drying.}
\label{fig:large_particles}
\end{center}
\end{figure*}

First and foremost, we demonstrate that the Cassie deposits are only observed for particles smaller than the top size of the pillars. Figure \ref{fig:large_particles}-\textit{Left} shows the final deposit of a evaporated drop containing 10 $\mu$m particles - thus of the same diameter as the pillars - at a (relatively high) initial concentration of 1 \% in mass: both the global image and local careful observations show that all the particles lie within a single central deposit, which for this high concentration take the shape of a dome. For lower concentrations (smaller than 0.1 \%), a more usual circular ring is observed, as there is not enough particles to be collected into a dome.

Other attempts with slightly smaller particles (5 to 10 $\mu$m) revealed the same trend. With these smaller particles, Cassie deposits consisting of single particles can appear for high initial concentration, near the Wenzel deposit. This is illustrated in Fig.~\ref{fig:large_particles}-\textit{Right}, where a few yeast cells - of typical size ranging from 5.5 to 8 $\mu$m - can stay on top of the posts. This image was obtained for a rather high concentration of yeast (evaluated to 0.1 \% in mass), but most cells lie within the grooves between posts after drying. The negligible percentage of cells lying on top of posts makes these surfaces inadequate for localization of particles of this range of size.


\subsection{Drops of colloidal particles suspensions} 

We now use drops of suspensions with 1 $\mu$m particles. The general aspect of the deposit is pictured out in Fig.~\ref{fig:depot_large}: the central Wenzel deposit, where most particles are collected - especially at high concentration - is generally 1.5 to 5 times smaller (in average radius) than the area encompassing the Cassie deposits. Quantitative and systematic observations with other particle sizes showed this general feature appears for any particle size, providing the diameter $d_p$ is much smaller than the size of the pillars: $d_p \ll d$. For easier observations, we decided to present results with 1 $\mu$m particles which, thanks to their size and relative brightness on fluorescence, enabled for the best contrast.

\begin{figure*}
\begin{center}
\includegraphics[scale=0.3]{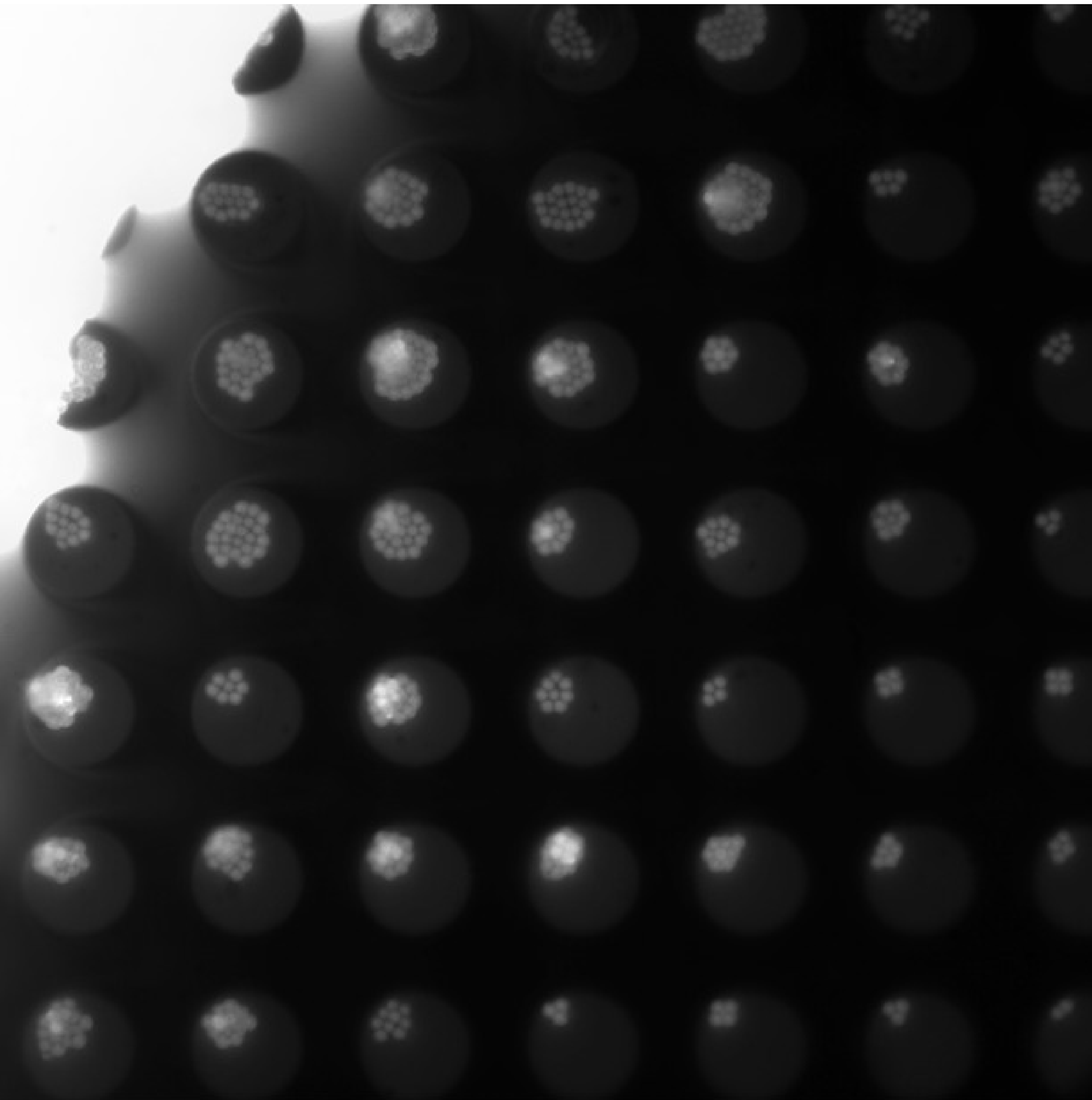}
\includegraphics[scale=0.3]{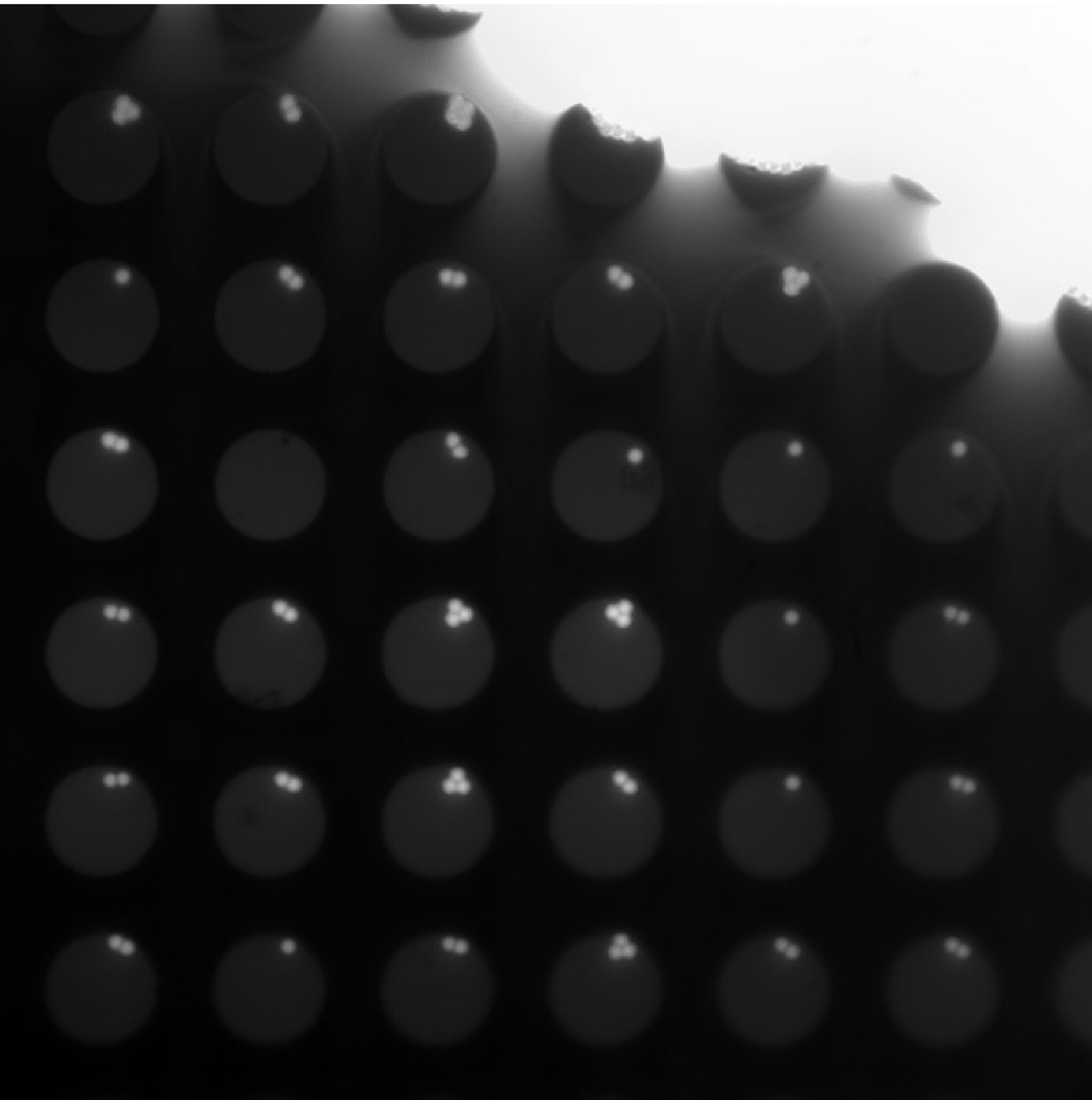}
\caption{Extracts at the edge of a deposit after the evaporation of a 10 $\mu$l drop of a suspension of 1 $\mu$m particles at concentration $c$ = 0.1 \%. \textit{Left} - In the vicinity of a preferential line of deposition, where deposit volumes can be rather large. \textit{Right} - Far from a preferential line of deposition, deposits can be quite small, even near the main ring.}
\label{fig:extract_fluo_150X}
\end{center}
\end{figure*}

The focus of our interest here is the spatial distribution of the size of the Cassie deposits. By varying different parameters like the initial concentration $c_0$ and volume $V_0$, the following trends have been noticed:

(1) the average size of individual Cassie deposits increases as the drop evaporates and recedes: larger Cassie deposits are generally observed close to the edge of the Wenzel deposit. 

(2) as expected, the Cassie radius $R_C$ increases with $V_0$, but also $R_C$ increases for higher $c_0$. It is only for high concentration that $R_C$ equals $R_0$ the initial base radius of the drop, otherwise $R_C \le R_0$. Therefore, the particles deposition does not generally occur in the first steps of the evaporation.

(3) the Cassie deposits are localized at the edge of the posts, and in angular positions corresponding to the receding direction (see left insert in Fig.~\ref{fig:depot_large} and Figs.~\ref{fig:extract_fluo_150X}).

\begin{figure}
\begin{center}
\includegraphics[scale=0.27]{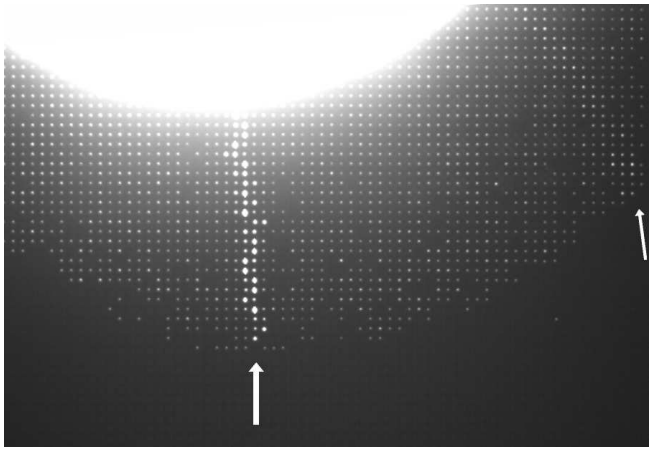}
\caption{The trailing edge of an evaporating drop, evidencing preferential directions with larger deposits (indicated by arrows).}
\label{fig:trace_drop_evap}
\end{center}
\end{figure}

(4) there exists preferential directions along which larger volumes of particles are deposited. This was already suggested in Fig.~\ref{fig:depot_large}, but it is even clearer in Fig.~\ref{fig:trace_drop_evap}. A careful examination of the position and shape of the drop during evaporation reveals that the tracks of these preferential deposition correspond to situations where the contact-line receded normal to the main axes of the network of pillars.

\section{Quantitative results}

\subsection{Influence of initial concentration}

\begin{figure}
\begin{center}
\includegraphics[scale=0.35]{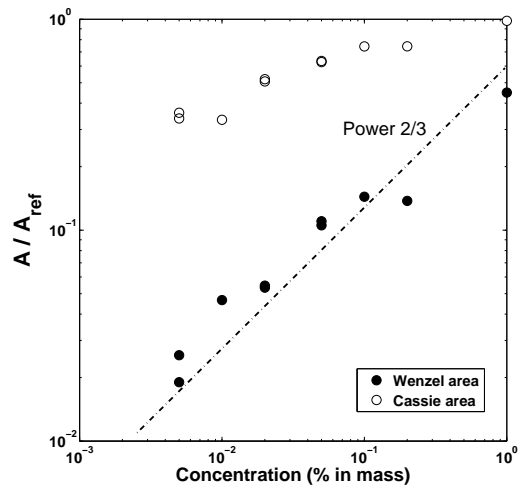}
\caption{Area of the Cassie and Wenzel deposits (divided by the relative base area of the initial drop) versus initial concentration $c_0$. Drop volume is 10 $\mu$l.}
\label{fig:area_vs_c_adim}
\end{center}
\end{figure}

Figure \ref{fig:area_vs_c_adim} shows the Cassie and Wenzel areas, made dimensionless via a division by the initial base area of the drop $A_{\text{ref}} = \pi r_0^2$, versus $c_0$ for drops of initial volume $V_0$ = 10 $\mu$l. Let us recall that the surface of the Wenzel area is directly governed by the critical drop radius $R_c$ for which the capillary pressure induces impalement. The area is simply equal to  $A = \pi r_c^2 = \pi R_c^2 (\sin \theta_r)^2$. Therefore, Fig.~\ref{fig:area_vs_c_adim} immediately gives the radius for liquid impalement versus $c_0$.

It is remarkable that the Wenzel area is significantly influenced by $c_0$. While for small concentration the critical base radius $r_c$ is much smaller than $r_0$, by a factor of about 10, at high concentration ($c_0 =$ 1 \%) $r_c$ is about one half of the initial radius (Fig.~\ref{fig:area_vs_c_adim}). The plot suggests that the Wenzel area increases with $c_0$ with the following power law: 

$$ \frac{A}{A_{\text{ref}}} \sim c_0^{2/3} $$

What does this trend suggest ? Let us evaluate $c_c$, defined as the concentration of particles when impalement occurs. The radius $R$ of the spherical cap intersecting a drop of volume $V$ with contact-angle $\theta$ is set by:

$$ R = \left( \frac{3 V}{\pi (2+ \cos \theta)(1- \cos \theta)^2} \right)^{1/3} $$

Therefore, if we consider than the evaporating drop recedes on the surface with an angle $\theta_r$ = 125$^{\circ}$, the base radius of the drop $r_0$ reads:

\begin{equation}
r_0 = R_0 \mid \sin \theta_r \mid = \left( \frac{3 V  \mid \sin \theta_r \mid^3}{\pi (2+ \cos \theta_r)(1- \cos \theta_r)^2} \right)^{1/3}
\label{eq:radius_base}
\end{equation}

\noindent and this equality is verified all along the evaporation process replacing $r_0$ and $R_0$ by $r(t)$ and $R(t)$.

Taking into account the very small amount of particles in Cassie deposits, we assume that the loss of particles deposited on top of the pillars during the first phase of evaporation - before the Cassie-to-Wenzel transition - is negligible. The quantity of particles remaining in the drop when impalement occurs is then about the same as initially: $N_0 = V_0 \times c_0 = V_c \times c_c$. Therefore, the critical concentration is:

\begin{equation}
c_c = \frac{3 \mid \sin \theta_r \mid^3}{\pi (2+ \cos \theta_r)(1- \cos \theta_r)^2} c_0 V_0 \frac{1}{r_c^3}
\label{eq:c_vs_t}
\end{equation}

Considering the trend shown in Fig.~\ref{fig:area_vs_c_adim}, the critical radius scales as $r_c \simeq \alpha \times c_0^{1/3}$, where $\alpha$ is an empirical constant of the right dimension. Together with eq.~(\ref{eq:c_vs_t}), this means that the concentration $c_c$ is independent of $c_0$, and then in all the experiments of Fig.~\ref{fig:area_vs_c_adim} the liquid impalement occurs \textit{at constant concentration}. Clearly in these experiments, it implies that impalement does not occur at constant drop radius. This result is at odds to what previous experiments of evaporation of pure water (without particles) droplet on SH showed \cite{Bartolo06,Reyssat08,Tsai10}, where the critical drop radius at impalement mainly depended on the pillar and network geometry. Therefore, contrary to the experiments with pure water, the liquid impalement of suspensions is not \textit{only} provoked by capillary pressure.

A first naive explanation is that the colloidal particles would act as surfactants, which would tend to favor impalement by decreasing the surface energy of the Wenzel state. However, this is dismissed by that the apparent contact-angle of particle laden drops is independent of $c_0$ even at high concentration: we did not notice any change of the apparent receding contact-angle $\theta_r$, while a decrease in surface tension $\sigma$ due to particles would have modified $\theta_r$. Another possible explanation is that once the concentration $c_(t)$ has reached a certain upper threshold, the capillary pressure is not simply given only by the drop radius: the accumulation of particles at the contact-line, and the subsequent increasing pinning force, may also enhance the liquid impalement. This interpretation can be supported by predictions and observations of an increased self-pinning induced by solute confinement at the contact-line \cite{Sangani09,Chhasatia11}. Therefore, an increasingly strong pinning at higher $c(t)$ would lead to higher local bending of the interface, and in turn to liquid impalement. Furthermore, the large deposits covering the whole top of posts and occurring just before Cassie-Wenzel transition might enhance the liquid impalement by smoothing the (initially sharp) shape of posts. 

The Cassie area seems to be less influenced by initial concentration than the Wenzel area, although a monotonic increase of $A_{\text{Cassie}}$ with $c_0$ is clearly visible in Fig.~\ref{fig:area_vs_c_adim}. From what precedes, one easily deduces that the occurrence of the first Cassie deposit during evaporation does not generally correspond to the initial base radius ($r_c \le r_0$), except for the high concentration $c_0$ = 1 \%. Otherwise at lower concentration, the drop starts to recede without leaving any particles on top of pillars.

\subsection{Influence of drop initial volume}

\begin{figure}
\begin{center}
\includegraphics[scale=0.40]{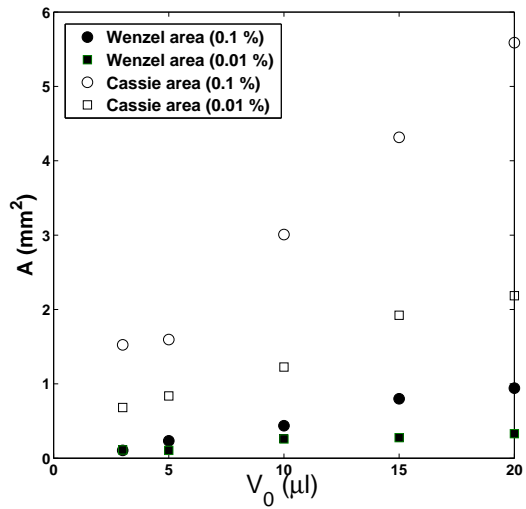}
\caption{Area of the Cassie and Wenzel deposits in dimensional units, versus volume for two initial concentration.}
\label{fig:area_vs_volume_1um}
\end{center}
\end{figure}

From a given $c_0$, the initial volume influences the amount of solid material suspended inside the drop. In the range of volumes used here (3 to 20 $\mu$l), the drop is never immediately impaled inside the texture. In other terms, and similarly to the previous series of experiments, the initial capillary pressure exerted by the drop $P_c = \frac{2 \gamma}{R_c} $ is below the critical pressure for impalement. The Wenzel area $A_{\text{Wenzel}}$ is then smaller than the initial drop base surface $A_{\text{ref}}$. Figure \ref{fig:area_vs_volume_1um} shows the Cassie and Wenzel areas versus $V_0$, for two initial concentrations $c_0$ (0.01 \% and 0.1 \%). Not very surprisingly, the Cassie and Wenzel areas increase with initial volume, and these area are always bigger for $c_0$=0.1 \% than for $c_0$=0.01 \%. This is consistent with the previous results in Fig.~\ref{fig:area_vs_c_adim}: both plots show that the size (base radius) of the Wenzel deposit increases with the \textit{initial amount of particles} $N_0$ and hence, that the liquid impalement does not occur at constant capillary pressure. In a similar fashion, the size of the Cassie-deposit area increases with $N_0$.

\begin{figure}
\begin{center}
\includegraphics[scale=0.45]{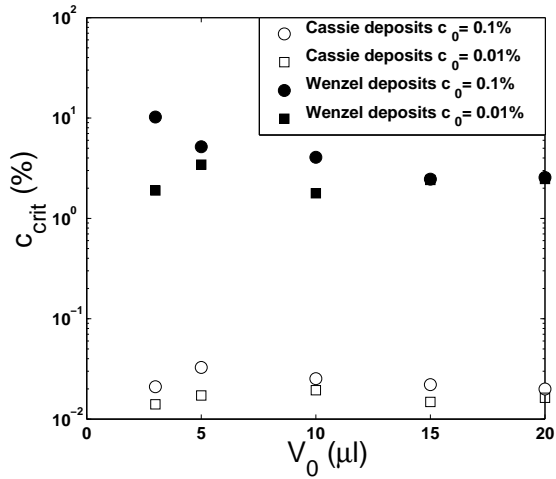}
\caption{Critical concentrations based on the average radius of Cassie and Wenzel deposits.}
\label{fig:ccrit_vs_vol}
\end{center}
\end{figure}

This suggest that the occurrence of the first Cassie deposits could occur once the concentration of particles in the drop has reached a threshold during evaporation. To check this, we plotted the critical concentration $c_c$ corresponding to the occurrence of the most peripheral deposit versus initial volume (see Fig.~\ref{fig:ccrit_vs_vol}), for two different $c_0$. The graph also includes the critical concentration for the Cassie-Wenzel transition. This reveals that the first Cassie deposit occurs for a concentration between $0.015$ and $0.03$ $\%$ in mass.

\section{On the origin of the Cassie deposits}

The unexpected appearance of deposits formed during the retraction phase in a Cassie state has recently found a convincing explanation. Indeed, similar tiny deposits have been noticed after the sliding of drops on such micro-textured superhydrophobic surfaces: using a UV-curable resist, Dufour \textit{et al.} were able to monitor tiny amounts of this non-volatile liquid pinned on top of each posts crossed by the sliding droplet  \cite{Dufour12}. The UV-exposure then allowed to observe the frozen interface with a SEM. As a counterpart of the small surface tension of this resist ($\gamma$=0.04 N/m), the posts were designed to have an overhanging structure: indeed, this geometry allows for the repelling of low surface tension liquids \cite{Tuteja08} (super-oleophobicity). On usual straight posts, the UV-curable resist gets immediately impaled. It is to be noted that, in order to check the universal character of such observations, similar experiments were carried out with non-volatile glycerin - of surface tension $\gamma$ = 0.065N/m slightly smaller than that of water - on straight posts. Although this universal behavior was indeed remarked in post-sliding visualizations, the use of liquid glycerin could not allow for SEM observations.

\begin{figure*}
\begin{center}
(a)\includegraphics[scale=0.32]{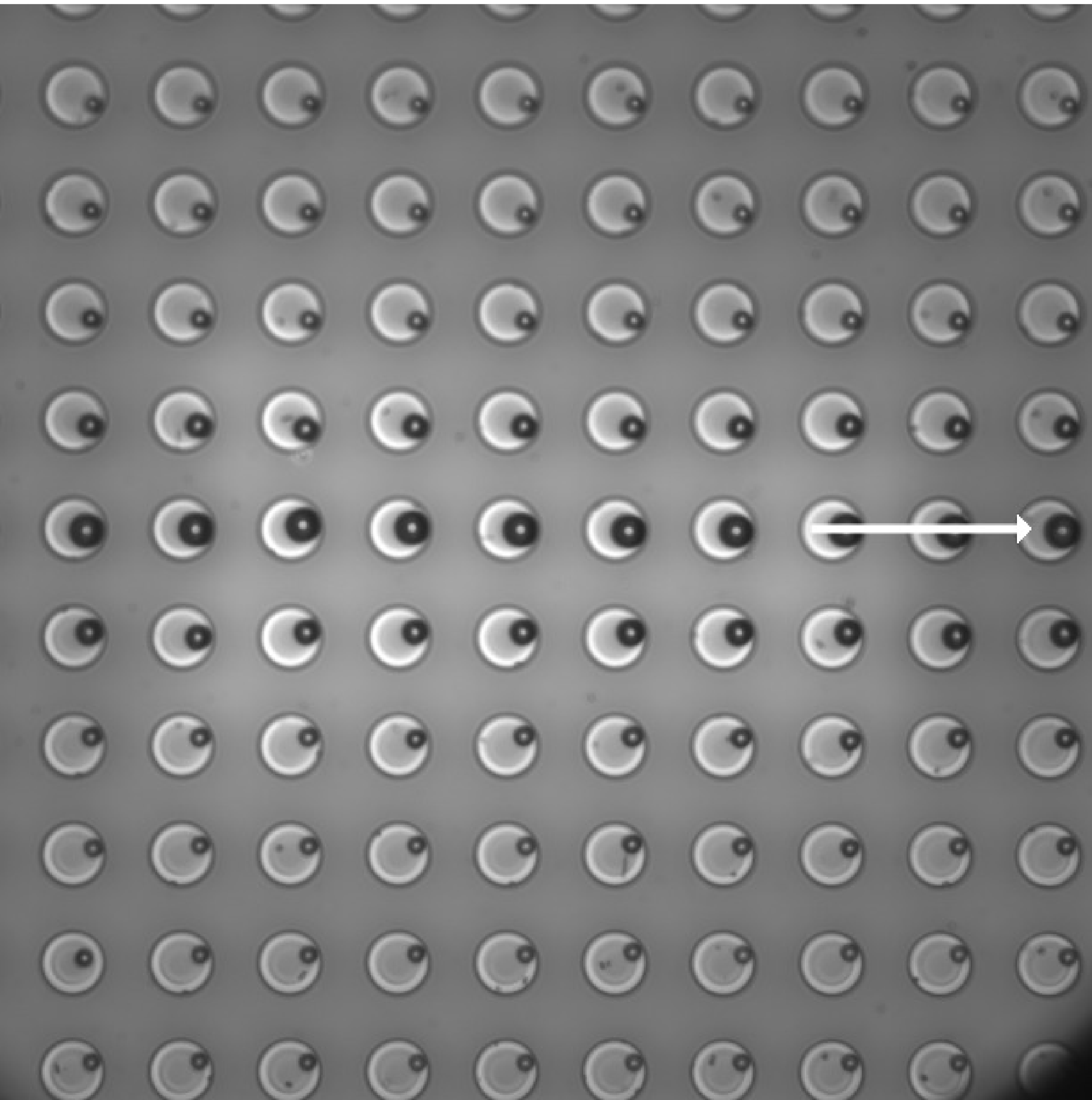}
(b)\includegraphics[scale=0.28]{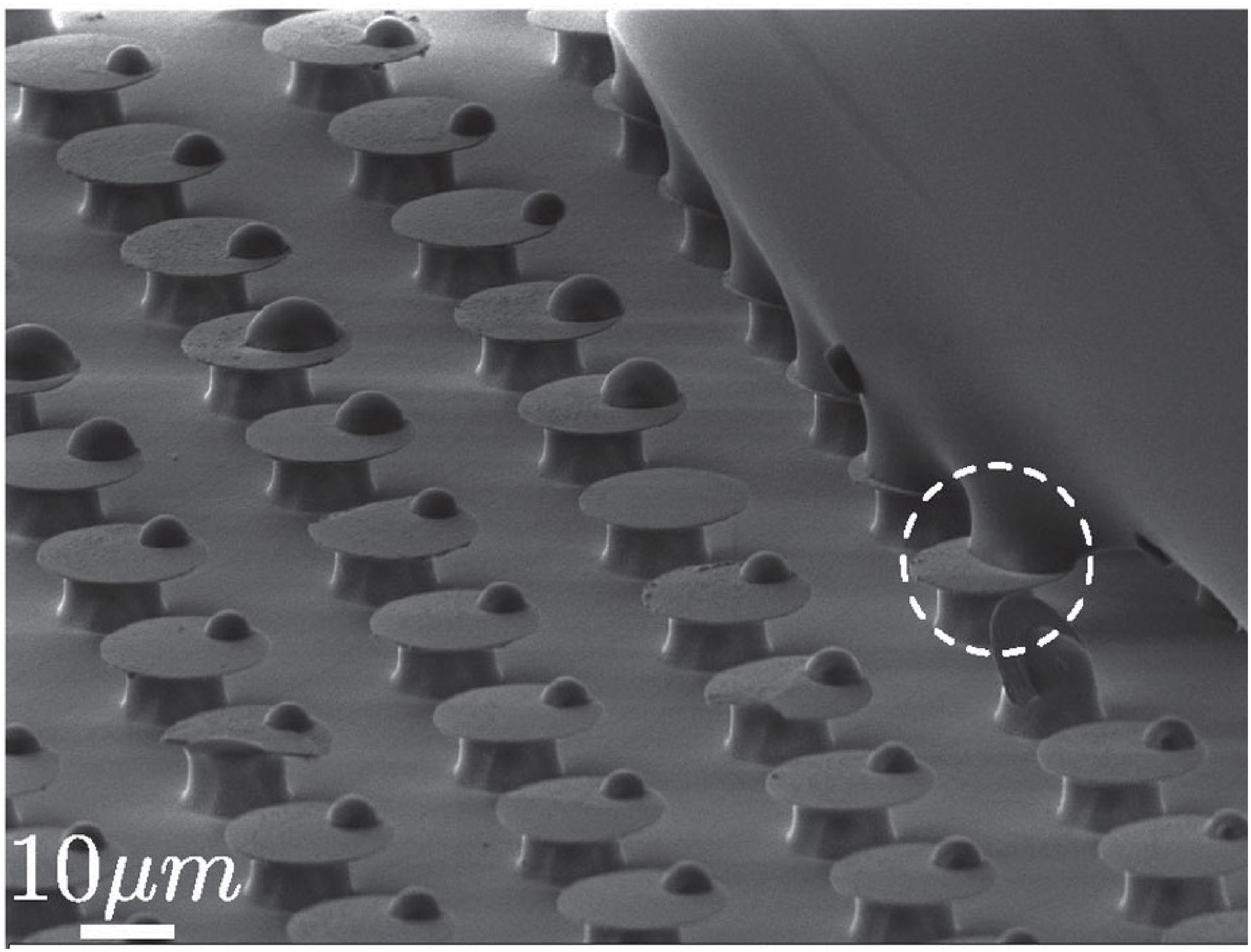}
\caption{\textit{Left} - An array of tiny droplets left on each posts (diameter ad spacing of 10 $\mu$m) after the sliding of a drop (direction indicated witht eh white arrow) of NDA curable resist. The biggest droplets lie along the row corresponding to the central axis of the drop. \textit{Right} - Sliding and zipping of the liquid interface on the array of posts. The white circle emphasizes an elementary sliding event.} 
\label{fig:goutte_NDA}
\end{center}
\end{figure*}

Figure \ref{fig:goutte_NDA}-(a) (obtained with an optical microscope) illustrates how tiny droplets get distributed at the trailing edge of a sliding drop. The axis of the drop motion corresponds to the occurrence of the biggest drops, which base surface occupies about 25 \% of the post total surface. The fine structure of the droplet volume spatial distribution also appears in evaporation experiments, as emphasized in Figure \ref{fig:trace_drop_evap}. Quantitative experiments in Dufour et al.'s paper \cite{Dufour12} also presented results on the dependance on sliding velocity of this spatial distribution: basically, a higher velocity promotes larger volumes for deposits and a smoother distribution. Evaporation experiments reported here, show rather sharp distribution of volumes, which is consistent with the fact that evaporation occurs at very small receding velocity. In this sense, it is to be noted that liquid deposition on to of posts of SH surfaces was already reported in \cite{Krumpfer11}, but with a homogeneous distribution of volumes (droplets base area were that of a post). 

Even more insightful is Figure \ref{fig:goutte_NDA}-(b), which offers a tridimensional image of the interface during drop sliding. This image shows that drop sliding and dewetting occurs by means of micro-capillary bridges formation. The liquidÐvapor interface at the drop trailing edge deforms until one reaches the receding contact angle $\theta_r$ on the defect boundary (Figure 3 a,b). During this process, several liquid bridges stretch. In the meantime, a dynamic phenomenon starts involving slip of the liquid across the structure cap over a length comprised between 0 and $d$, while the interface between two pillars rises to keep constant capillary pressure, until pinch-off occurs. Subsequently, this leaves a small amount of liquid on each pillar. 

These SEM shots offer a qualitative picture of the interface deformation and hydrodynamics at the scale of posts. Liquid drop motion on SH surfaces has been a major subject of investigations over the last decade, and such liquid pinning is not generally taken into account in the determination of hysteresis \cite{Dorrer_Ruhe_06}. Though, more recent studies suggest that capillary bridge stretching and pinch-off on each post - considered as 'strong defect' - should significantly contribute to CA hysteresis and retention forces \cite{Reyssat09}. Of course, such microscopic deformations should be addressed theoretically with more care (see eg. Dupuis and Yeomans' paper \cite{Dupuis_Yeomans06}), but our drying solute-in-drops experiments confirm the importance of capillary bridge formation and break-up in such processes. 

\section{Conclusions}

The drying of colloidal suspensions on SH surfaces exhibits a complex structure of deposits which, additionally to the classical "coffee-ring", shows tiny deposits on top of the posts outwards the main ring. The latter is obtained once the drop has been impaled into the texture and gets strongly pinned at its contact-line, while tiny deposits occur while the drop still sits on top of pillars in a Cassie-Baxter state. As revealed by experiments with a non volatile and UV-curable liquid, the presence of these deposits is the direct consequence of a weak but finite pinning force on top of each pillar, which leads to the retention of small liquid droplets and after drying of these droplets, leads to the deposition of the inside colloidal particles. Therefore, the volume distribution of these deposits is ruled by complex hydrodynamics processes: local pinning of contact-line, stretching, pinch-off of a liquid bridge and detachment of liquid. Additionally to these effects, the presence of particles at relatively high concentration shall influence the pinning force on each post.

From these experiments, we can draw important consequences about which type of - and under which conditions - SH surfaces can be qualified as "self-cleaning". Conversely, other types of SH surfaces should be more suitable for solute targeting or pre-concentration of suspensions. We can summarize in several points what these experimental results suggest in such applied prospectives:

- The first obvious point is that the robustness of the surface against liquid impalement, quantified by the critical liquid pressure required to undergo Cassie-to-Wenzel transition, has to be as strong as possible. Ideally, the impalement transition should not occur at all during drying. This should be the case if the liquid droplets used for cleaning are large enough or if the posts are thin and tall enough \cite{Bartolo06,Reyssat06,Lapierre10}. This situation can especially be suitable for the pre-concentration of solutes. In this sense, a very recent study \cite{Alvaro12} reported on the capillary assembling of colloidal particles into a single spherical deposit. However, although the drop would stay in a Cassie state during drying, it does not necessarily imply that the top of the structures would not be covered by tiny particle deposits.

- Although this specific point would require further examination and complementary experiments, the self-cleaning potential of super-hydrophobic surfaces seems to depend on the relative size of the dirt particles with respect to the top surface of the structures (this is explicitly stated in Fig. \ref{fig:large_particles}. If one knows in advance that the particles to remove and transport out of the surface are big enough, there should be no need to design a textured surface with thin posts. This is directly the consequence of that any droplet stuck on top of a post must have a larger volume than that of particles in order to leave at least one on a post after evaporation.  

- For solute targeting (tiny amount of particles on specific locations), it seems better to have large and sharp-shaped posts, which should ensure liquid pinning. The volume of droplets is strongly influenced by sliding velocity and relative angle between sliding direction and eigen axes of the network of posts.

\subsection*{Acknowledgements}

Y. Coffinier and T.P.N. Nguyen are kindly acknowledged for having provided the micro-textured surfaces. The interpretation of results benefited from fruitful discussions with R. Dufour. The fluorescence visualizations benefited from the valuable help of F. Zoueshtiagh.


\begin{thebibliography}{100} 

\bibitem{Plawsky08}
J.L. Plawsky, M. Ojha, A. Chatterjee, P.C. Wayner, Jr,
\textit{Chem. Eng. Comm.}, 2008, \textbf{196,} 658-696.

\bibitem{Deegan97}
R.D. Deegan \textit{et al.},
\textit{Nature}, 1997, \textbf{389,} 827-829.

\bibitem{Nellimoottil07}
T.T. Nellimoottil, P.N. Rao, S.S. Ghosh, A. Chattopadhyay,
\textit{Langmuir}, 2007, \textbf{23,} 8655-8658.

\bibitem{Baughman10}
K.F. Baughman \textit{et al.}
\textit{Langmuir}, 2010, \textbf{26,} 7293-7298.

\bibitem{Hu_Larson06}
H. Hu and R.G. Larson,
\textit{J. Phys. Chem. B}, 2006, \textbf{110,} 7090.

\bibitem{Ristenpart07}
W.D. Ristenpart, P.G. Kim, C. Domingues, J. Wan and H.A. Stone,
\textit{Phys. Rev. Lett.}, 2007, \textbf{99,} 23450.

\bibitem{Rio2006}
E. Rio, A. Daerr, F. Lequeux and L. Limat,
\textit{Langmuir}, 2006, \textbf{22,} 3186-3191.

\bibitem{Bodiguel2010}
H. Bodiguel, F. Doumenc and B. Guerrier,
\textit{Langmuir}, 2010, \textbf{26,} 10758-10763.

\bibitem{Berteloot12}
G. Berteloot, A. Hoang, A. Daerr, H. P. Kavehpour, F. Lequeux, L. Limat,
\textit{J. Coll. Interf. Sci.}, 2012, \textbf{370,} 155-161.

\bibitem{Narayanan04}
S. Narayanan, J. Wang and X-M. Lin,
\textit{Phys. Rev. Lett.}, 2004, \textbf{93,} 135503.

\bibitem{Ressine08}
A. Ressine, D. Finnskog, G. Marko-Varga and T. Laurell,
\textit{Nanobiotechnol.}, 2008, \textbf{4,} 18Ð27.

\bibitem{Sangani09}
A.S. Sangani, C. Lu, K. Su and J.A. Schwarz,
\textit{Phys. Rev. E.}, 2009, \textbf{80,} 011603.

\bibitem{Onda96}
T. Onda, S. Shibuichi, N. Satoh and K. Tsujii,
\textit{Langmuir}, 1996, \textbf{12,} 2125.

\bibitem{Cassie}
A. Cassie and S. Baxter,
\textit{Trans. Faraday Soc.}, 1944, \textbf{40,} 546.

\bibitem{Bico99}
J. Bico, C. Marzolin and D. Qu\'er\'e,
\textit{Europhys. Lett.}, 1999, \textbf{47,} 220-226.

\bibitem{Tsai10}
P. Tsai, R.G.H. Lammertink, M. Wessling and L. Lohse,
\textit{Phys. Rev. Lett.}, 2010, \textbf{104,} 116102.

\bibitem{McHale05}
G. McHale, S. Aqil, N.J. Shirtcliffe, M.I. Newton and H.Y. Herbil,
\textit{Langmuir}, 2005, \textbf{21,} 11053-11060.

\bibitem{Barthlott97}
W. Barthlott and N. Neinhuis,
\textit{Planta}, 1997, \textbf{202,} 1-8.

\bibitem{GaoJiang04}
X. Gao and L. Jiang,
\textit{Nature}, 2004, \textbf{432,} 36.

\bibitem{Bartolo06}
D. Bartolo, F. Bouamrirene, E. Verneuil, A. Buguin, P. Silberzan and S. Moulinet,
\textit{Europhys. Lett.}, 2006, \textbf{74,} 299.

\bibitem{Reyssat06}
M. Reyssat, A. Pepin, F. Marty, Y. Chen, and Qu\'er\'e,
\textit{Europhys. Lett.}, 2006, \textbf{74,} 306.

\bibitem{Lapierre10}
F. Lapierre, P. Brunet, Y. Coffinier, V. Thomy, R. Blossey and R. Boukherroub,
\textit{Faraday Disc.}, 2010, \textbf{146,} 125-139.

\bibitem{Nguyen10}
T.P.N. Nguyen, P. Brunet, Y. Coffinier and R. Boukherroub,
\textit{Langmuir}, 2010, \textbf{26,} 18369Ð18373.

\bibitem{Neto_etal11}
A.I. Neto, C.A. Custodio, W. Song and J.F. Mano,
\textit{Soft Matt.}, 2011, \textbf{7,} 4147-4151.

\bibitem{Shao_etal12}
F. Shao, T.W. Ng, O.W. Liew, J. Fu and T. Sridhar,
\textit{Soft Matt.}, 2012, \textit{DOI: 10.1039/c2sm07127d}.

\bibitem{Wenzel}
R. Wenzel,
\textit{Ind. Eng. Chem.}, 1936, \textbf{28,} 988.

\bibitem{Herminghaus00}
S. Herminghaus,
\textit{Europhys. Lett.}, 2000, \textbf{52,} 165-170.

\bibitem{Reyssat08}
M. Reyssat, J.M. Yeomans and D. Qu\'er\'e,
\textit{EPL}, 2008, \textbf{81,} 26006.

\bibitem{Chhasatia11}
V.H. Chhasatia and Y. Sun,
\textit{Soft Matter}, 2011, \textbf{7,} 10135.

\bibitem{Dufour12}
R. Dufour, P. Brunet, M. Harnois, V. Thomy, R. Boukherroub and V. Senez,
\textit{Small}, 2012, \textbf{8,} 1229-1236.

\bibitem{Tuteja08}
A. Tuteja, W. Choi, J.M.Mabry, G.H. McKinley and R.E. Cohen,
\textit{Proc. Natl. Acad. Sci.}, 2008, \textbf{105}, 18200-18205.

\bibitem{Krumpfer11}
J.W. Krumpfer, P. Bian, P. Zheng, L. Gao and T.J. McCarthy,
\textit{Langmuir}, 2011, \textbf{27,} 2166Ð2169.

\bibitem{Dorrer_Ruhe_06} 
C. Dorrer and J. Ruhe,
\textit{Langmuir}, 2006, \textbf{22,} 7652-7657.

\bibitem{Reyssat09}
M. Reyssat and D. Qu\'er\'e,
\textit{J. Phys. Chem. B}, 2009, \textbf{113,} 3906-3909.

\bibitem{Dufour11}
R. Dufour, M. Harnois, V. Thomy, R. Boukherroub and V. Senez, 
\textit{Soft Matt.}, 2011, \textbf{7,} 9380-9387.

\bibitem{Dupuis_Yeomans06}
A. Dupuis and J.M. Yeomans,
\textit{Europhys. Lett.}, 2006, \textbf{75,} 105-111.

\bibitem{Alvaro12}
A.G. Marin \textit{et al.},
arXiv:1203.4361v1 [cond-mat.soft] (2012).


\end{thebibliography}
\end{document}